# Improving Usability of Interactive Graphics Specification and Implementation with Picking Views and Inverse Transformations

Stéphane Conversy
University of Toulouse – ENAC - IRIT
Toulouse, France
stephane.conversy@enac.fr

*Abstract*— **Specifying and programming graphical interactions are difficult tasks, notably because designers have difficulties to express the dynamics of the interaction. This paper shows how the MDPC architecture improves the usability of the specification and the implementation of graphical interaction. The architecture is based on the use of picking views and inverse transforms from the graphics to the data. With three examples of graphical interaction, we show how to express them with the architecture, how to implement them, and how this improves programming usability. Moreover, we show that it enables implementing graphical interaction without a scene graph. This kind of code prevents from errors due to cache consistency management.**

## I. INTRODUCTION

Interactive system programming is difficult, notably because designers have difficulties to express the dynamics of the interaction [1]. Even if interaction is inherently graphical, specifying it and implementing it still relies mainly on textual languages that enlarge the gap between the phenomenon to describe and the description. Furthermore, writing interactive code with calculus-oriented languages are not suitable to describing reactive processes [2][3]. This results in so-called spaghetti [2] code that prevents readability, and that favors the emergence of bugs, notably when the system grows after several increments. Finally, the need to make systems as fast as required by the interaction loop (short duration between user action/machine reaction/user perception) forces the designers to optimize their code, and to make it difficult to read or modify.

We think that these problems pertain to the usability of specification and implementation of interactive graphics. Specifying interaction consists in describing how graphical representations react to user input (referred as "designing" in [1]). This is a problem that has been approached before with various languages (including visual), but as noted in [1], further work needs to be done to facilitate this task. Implementation pertains to the process with which a programmer can turn a specification into executable code. Again, various approaches aimed at improving the transition, and the readability of interaction code. However, we think that a number of unimportant considerations hinder readability of code, and that a better architecture is necessary.

In this paper, we rely on a particular architecture to ease specifying interactive graphics, and to ease implementation of interactive graphics. The specification narrows the gap between the phenomenon and its description. The implementation paradigm enables to use a data-flow architecture, which is more readable and more manageable than imperative code. We first present the pattern on which this work relies. After discussing a number of dimensions of analysis, we then show three examples of interactive graphics, and discuss why we think their specification and their implementation is more readable and understandable.

## II. MDPC

This section briefly introduces MDPC, the pattern we used (more details are available in [4]). MDPC (Model – Display view – Picking View – Controller) relies on two principles: "picking views", and "graphical transformations". Picking views are invisible graphical objects that reify spatial modes of interaction. A spatial mode is the spatial equivalent of a temporal mode: different behavior in function of space, versus different behavior in function of time. Figure 1. shows the "display view" of a hierarchical menu (top left), and the corresponding picking view when the user is navigating in the menu (top right). The (transient) triangle laid over the menu in the picking view enables to reach the sub-menu entries while avoiding submenu folding. Picking views have two benefits. First, they help managing the dynamic of the states of the interaction (here the transient triangle), as opposed to the graphical state of the display. Second, they enable to avoid analytical computation of spatial relationships (here a movement with a direction below 45°) by using Enter/Leave events generated by the underlying graphical toolkit.

Graphical transformations are functions that transform the conceptual model into graphics. MDPC uses two graphical transformations: one for the display view, and one for the picking view. Figure 1. shows the affine transforms applied to the model (two values between 0 and 1) to generate the display view and the picking view of a horizontal scrollbar. Computing the inverse transformations enable to translate a graphical interaction (say a drag of the thumb) into operations on the model (translation of both values).

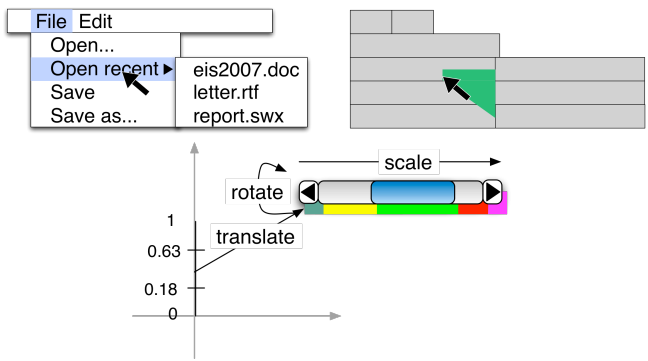

Figure 1. The display view of a menu (top left) and its picking view (top right). The display and picking view (shifted for clarity) of a scrollbar, and the transforms from the model to the views (bottom).

MVC was the result of the application of the separation of concerns principle on interactive code [5] that aims at improving modularity. MDPC [4] is a refinement of the MVC architecture (hence the name MDPC: the View in MVC is split in D and P, Display View and Picking view), and can be considered as the application of separation of concerns down to the MVC Controller itself (see Figure 2. ). By using a picking view and inverse transforms, MDPC offloads from the MVC Controller the management of interactive state, and the translation of events into operations on Model. This makes the Controller code much simpler, almost eliminates the apparent impossibility to decouple the Controller and the View, and makes Views and Controllers invariant from geometrical and layout transforms. This also improves modularity, since the Controller can be made more general and reusable. For example, the same Controller can be used for various species of scrollbar (arrows at both extremities, at each extremity, on the thumb; horizontal, vertical, and radial layout). MDPC has been shown to make possible entirely "model driven" implementation of scrollbars, sliders, range-sliders and hierarchical menus.

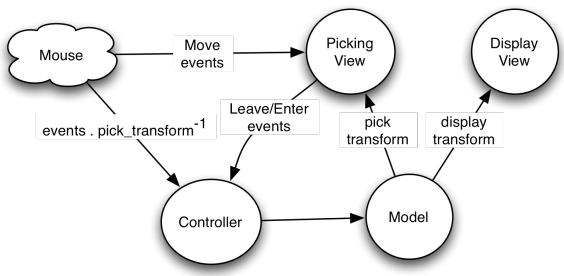

Figure 2. MDPC architecture

## III. DIMENSIONS OF ANALYSIS

We think that MDPC is also beneficial to the specification of interactive graphics, and to their implementation. More precisely, using MDPC as a pattern helps at both designing the specification, and designing the code. As such, MDPC can be considered as a method that improves the usability of programming.

Usability of "programming" (taken in the large, i.e including specification and implementation) is the extent to which an environment (including language, pattern, IDEs etc) can be used to achieve programming tasks with effectiveness, efficiency, and satisfaction (see [6] for on application on classes). Usability is difficult to assess, because it requires longitudinal studies with a large number of designers (as defined in [1]). Since we have not done such studies in this work, we provide predictive evaluation of specification usability and implementation usability along three properties.

The first property that we assess is the descriptive power. i.e. the extent to which a designer using MDPC is able to specify and implement existing graphical interactions. This is a prerequisite for designers if we want them to be effective: they will not be able to design an intended interaction if the architecture does not allow for it. In the next section, we present three examples of specification and implementation of interactive graphics: Drag'n'Drop with hysteresis (direct manipulation technique [7]), Magnetic guides (instrumental interaction technique [8]), and a Calendar (complex representation combined with direct manipulation). Together, those examples aimed at showing that MDPC expressive power is sufficient to specify a large range of graphical interactions. In addition, we describe two kinds of implementation, one based on a scene-graph (D'n'D, Magnetic guides), and the other one based on a data-flow (Calendar). We show code snippets to help explain the implementation, to convince the reader that the implementation actually exists and runs, and to help the reader replicate this work.

The second property that we assess is simplicity of description. Even if MDPC has a sufficient descriptive power, it would be useless if the description itself were cumbersome to specify and program. We provide an evaluation of simplicity of description by using concepts from the Cognitive Dimensions of Notation framework (CDN) [9], and from a list of desirable properties employed in the literature (see [3] for a survey).

The third property that we assess is the performance (implementation only). However elegant an implementation is, its usefulness can be reduced if performances are too weak. Hence, we also discuss this aspect.

## IV. DRAG'N'DROP WITH HYSTERESIS

The first example is the Drag'n'Drop with hysteresis, a direct manipulation technique. Drag'n'Drop with hysteresis forces the user to move passed a small minimum distance from the ButtonPress position, before effectively triggering the Drag operation. This prevents the system from misinterpreting a Selection for a Drag'n'Drop: when selecting a graphical object with a click (ButtonPress then ButtonRelease), one or a few « Move » events may occur between the button events, because the mouse slips due to the force applied by the pointing finger on the button. This makes the system misinterpret a Selection for a Drag'n'Drop, and moves the selected object by a slight but undesirable amount.

## A. Interaction specification

A traditional analytical algorithm consists in computing at each Move event the distance between the ButtonPress position and the cursor position, testing if the distance is superior to the minimum distance, and moving the object if the test is successful. This necessitates the computation of a Euclidean distance (square root of sum of squares).

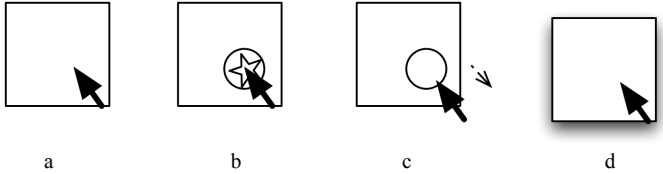

Figure 3. hysteresis with MDPC. (a) hover (b) press:an invisible circle is inserted into the scene (visible here for explanation) (c) no drag while the cursor stays in the circle (d) leaving circle: removal of the circle, drag starts.

### 1) Description

The version with MDPC consists in drawing an invisible circle centered on the position of the ButtonPress, with a radius equal to the hysteresis distance. Figure 3. shows the display and picking views for explanation purpose: the circle is visible, but in the real system it is not. At the beginning, the cursor is at the centre of the circle. If the cursor does not leave the circle before a ButtonRelease, the interaction is interpreted as a simple Select. If the cursor leaves the circle, the minimum distance is reached, and the Drag can start. The invisible circle is removed, which allows the user to move the object within a distance from its initial position smaller than the hysteresis distance.

### 2) Simplicity

We think that the MDPC description is closer to the conceptual model of the interaction. In fact, computing at each Move event the distance between the cursor and its initial position is not necessary to specify the interaction. The only needed information is the minimum distance to reach. Since the minimum distance to reach is reified into a circle, the concept of distance crossing is more directly represented. Hence, MDPC improves the Closeness of Mapping cognitive dimension. Finally, the designer can make visible picking view for debugging purpose. By directly seeing the circle on the screen, one can understand how the graphical interactive state behaves, and debugs more easily than with code only. Here, MDPC improves the Visibility cognitive dimension.

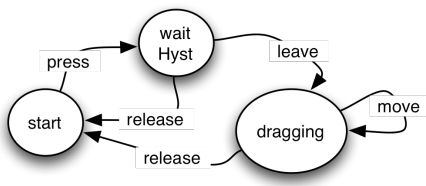

Figure 4. hysteresis state machine. Circles denote states, arrows transition. The text on a transition denotes the interaction event that triggers a transition.

## B. Implementation

This particular implementation uses the SwingStates toolkit [10]. SwingStates enables programming interaction with state machines directly in java files. The transition between states can be guarded (i.e. a predicate prevents the transition to fire), and be associated with an action when fired. SwingStates relies on a scene graph, i.e. a data structure that retains graphical objects. With SwingStates scene graph, graphical objects may be "tagged": any operation on an object can also be applied on a "tag", which means that any objects with this tag will be modified accordingly. We heavily used this feature in the following code.

```
public State start = new State() {
   Transition press = new PressOnShape(BUTTON1, ">> waitHyst") {
     public void action() {
        toMove = getShape(); // get the object to drag
        lastPoint = getPoint(); // store last clic position
        hystShape = new CEllipse(lastPoint.getX()-5,lastPoint.getY()-5, 10, 10); // picking shape
        hystShape.setDrawable(false); // set invisible
        canvas.addShape(hystShape); // add to scene graph
   }};};
```

Figure 5. action on "press" transition from "start" to "waitHyst".

### 1) Description

The state machine of Drag'n'Drop with Hysteresis is shown in Figure 4. When the user presses on an object, the current state becomes "waitHyst", and waits for the hysteresis distance to be crossed. The code of the action associated to the "Press" transition is shown in Figure 5. The picking shape is created (CEllipse, the circle), made invisible, then added to the scene graph. Graphical objects of the picking view are invisible to the user, but react to the mouse events. As said before, one can comment the line that make objects invisible for debugging purpose.

The circle is centered at the location of the cursor: hence, the cursor is inside the circle. The "leave" transition pertains to this circle (code not shown in the figure): when the cursor leaves the circle, the "leave" transition to the "Dragging" state is fired, an action removes the invisible circle from the scene graph, and the user is free to drag the object around.

### 2) Simplicity

Even if simple, the MDPC-based description of the interaction illustrates how Enter and Leave events are used in place of analytical computation of the Euclidean distance. Hence, the designer is not required to write this code. Of course, one can use abstraction in the code, and call a 'distance' function instead of the computation code, but the MDPC version gets rid of this necessity.

This example also illustrates how picking views help manage the dynamics of the interaction state. Finite State Machines are well adapted to MDPC descriptions of the interaction. At each state can correspond a particular picking view, which is active when the state is active. This is similar to the architecture described in [11]. Again, MDPC improves Closeness of Mapping with interactive state implementation.

### 3) Performance

Adding a single circle to a scene graph is inexpensive. The generation of Leave/Enter events may actually use a Euclidean distance, hence the computation is the same than the traditional algorithm.

## V. MAGNETIC GUIDES

Magnetic guides are instruments for aligning graphical objects [8]. During the Drag'n'Drop of an object, if the object is close enough to the magnetic guide, the guide attracts the object: hence, dropping multiple objects on a linear guide makes them aligned. More complex alignments allow for alignment of objects center, but also of their boundaries. More complex guides include Bezier curves.

Alignment with magnetic guides is an example of instrumental interaction [8]: a Magnetic Guide is an instance of an instrument i.e. action (alignment) reified into an interactive object that control other interactive objects. Magnetic guides are different from a "grid", since they are explicitly defined and manipulated by the user.

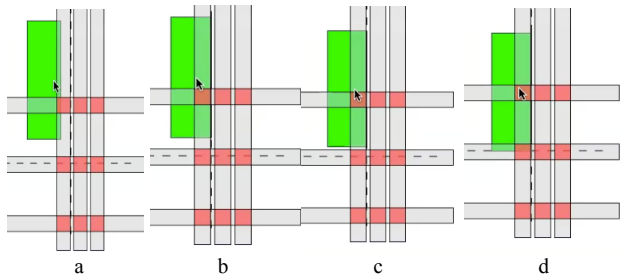

Figure 6. dashed lines: magnetic guides; gray rectangles : picking view of magnetic zones ; red squares: picking view of magnetic zones shared by two guides (a) free drag (b) right horizontal alignment (c) just before entering in the magnetic zone to align vertically, at the bottom (d)

### A. Interaction specification

Similarly to the previous example, a traditional analytical algorithm computes the distance between the guides and the dragged object, tests if the distance is inferior to the attraction distance, and sticks the object on the guide if so.

*1) Description*

Figure 6. shows both the display view (dashed line and green rectangle), and the picking view (gray rectangles, red square) for illustration purpose. With the MDPC pattern, the algorithm consists in drawing an invisible thick line over the guide (thin dashed line on the figure), which thickness is equal to the double of the attraction distance (Figure 6. , gray rectangles), and in registering a callback when the cursor enters or leaves the invisible thick line (events "Enter" and "Leave"). Thus, when the cursor enters the invisible thick line, the object sticks to the guide; when the cursor leaves the thick line, the object sticks to (and thus follows) the cursor.

As said earlier, more complex guides allow for alignment of objects center, but also of their boundaries. With MDPC, this is described with multiple picking zones, placed around the magnetic guides with respect to the geometry of the object, and the position of the cursor relative to the object (Figure 6. , gray rectangles).

Guides may intersect, which allows an object to stick to their intersection, and to preserve alignment with two sets of objects. Drawing two thick lines results in a partial occlusion of one line by the other at the intersection point. With a toolkit that can synthesize Enter and Leave events for occluded objects, no adaptation of the previous algorithm is necessary. However, with SwingStates' event synthesis model, the previous method does not work: the topmost guide would prevent the attraction from the occluded line since no Enter or Leave event would be emitted for the occluded thick line. To solve this problem, it is necessary to define the area of intersection between thick lines, and makes the object stick at the intersection when the cursor is in the intersection area (Figure 6. , red squares).

*2) Simplicity*

The interaction is complex, and the distances to compute are numerous: there are 6 distances per guide (3 vertical, 3 horizontal), and the reference point from which to compute the distance is not easy to grasp and understand. MDPC encourages the identification of spatial modes of interaction and their corresponding area. We think that thinking in terms of area of attraction is easier. As noted in [1], designers often use drawings to explore a solution, and explain them to colleagues. MDPC allows to directly using this drawings to express the interaction. In addition, when the guide themselves are complex (e.g. curves), no additional cost in terms of reasoning is necessary compared to the distance model. Similarly to the Drag'n'Drop example, MDPC thus improves Closeness of Mapping and Visibility.

The intersection area problem induces more coding for the designer than the distance computation model. The MDPC solution seems more complex than computing distances from guide: the burden of describing intersection shapes may no make MDPC as advantageous than claimed. This hinders the Terseness cognitive dimension. However, this problem only occurs with scene graphs that do not generate Enter/Leave events for occluded objects.

### B. Implementation

*1) Description*

This implementation also uses the SwingStates toolkit. The state machine is shown in Figure 7. The interaction begins with the hysteresis interaction described earlier. When crossing the hysteresis distance, the "leave" transition is fired, and the machine enters the "dragging" state.

Picking views are managed in the code of the action associated to the "leave" transition (shown in bold in Figure 7. The code itself is shown in Figure 8. First the previous picking views (hysteresis circle) is removed (a), and replaced by three picking objects per guide (b), to align according to the center and the boundaries of the dragged object. All picking objects are put on the position on the guide at first. The guides for the boundaries are then "spread around" the guide by a distance equal to half the height or width of the dragged object. Then, all guides are moved by a distance equal to the shift between the position of the cursor inside the dragged object, and its boundaries. Each time the cursor enters the picking shape of a guide (a thick line), the machine enters the corresponding state. In the "dragInXXGuide" state, the move transition triggers an action that moves the object along the guide. There is no "move" transition for the "inStickGuide", since no action is necessary.

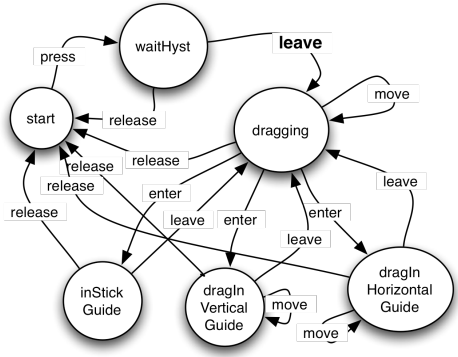

Figure 7. magnetic guide state machine

```
public State waitHyst = new State() {
  Transition drag = new LeaveOnShape(">> dragging") {
    public void action() {
      […]
      // (a) remove previous picking view
      canvas.removeShape(hystShape);
      // (b) create horizontal pick shapes
      for (int i=0; i<3; ++i) {
        CShape s = new CShape(new
BasicStroke(20).createStrokedShape(new Line2D.Double(0, 0, 500, 0)));
        canvas.addShape(s);
        s.addTag(hMagnetTag);
        // (c) spread the pick shapes around the guideline
        if (i==0) s.translateBy(0,toMove.getHeight()/2);
        if (i==2) s.translateBy(0,-toMove.getHeight()/2);
      }
      // translate around guideline
      hMagnetTag.translateBy(0,ymg);
      // (d) translate the pick shapes according to the relative position of
the cursor from the reference point of the shape (middle)
      hMagnetTag.translateBy(0,pickRelPos.getY());

      // create vertical pick shapes and sticky pick shapes at h and v
intersections
      // hidden: similar to horizontal guides
}}}
```

Figure 8. Action on "move" transition from "waitHyst" to "dragging"

*2) Simplicity*

For simple guides, such as horizontal or vertical guides, the computation of the position of the dragged object stuck to the guide is straightforward: one of the Cartesian dimensions is the one from the cursor, and the other is the one from the guide. In the case of a more complex guide, it is necessary to code the computation of the orthogonal projection of position of the dragged object on the guide, and sets its coordinates to the coordinates of the projection.

Since SwingStates does not synthesize Enter and Leave event for occluded objects, the code has to create the picking objects for the intersections. In the simple case of horizontal and vertical guides, the shape of the intersection is a square centered at the intersection of the guides. However, more complex guides may require more complex computation. In this case, MDPC extends nicely to the use of the AND operation of the constructive area geometry, and the computation the shape resulting from an AND between the two thick lines. Some toolkits provide such algorithms (i.e. Java2D Shape API, or OpenGL GLU tesselator [12]).

*3) Performance*

Again, in order to make reasoning easier, the code avoids analytical computation by relying on the algorithms provided by the scene graph. The test for shape belonging does not require a rasterization. Instead, the algorithm in the scene graph may use the distance algorithm that one would have used in the interaction code. Hence performances are similar.

With SwingStates model of events, additional computations of area are necessary. However, those computations happen only once during the interaction (in the transition between "waitHyst" and "Dragging").

## VI. CALENDAR

The next example is a Calendar application, with a "week" view on events, such as Apple's iCal or Google Agenda. We have replicated two interactions: "Drag'n'Drop" of calendar events, which allows to move an event in the day, or to move it into another day of the displayed week; and the "Resize" of the duration of calendar events.

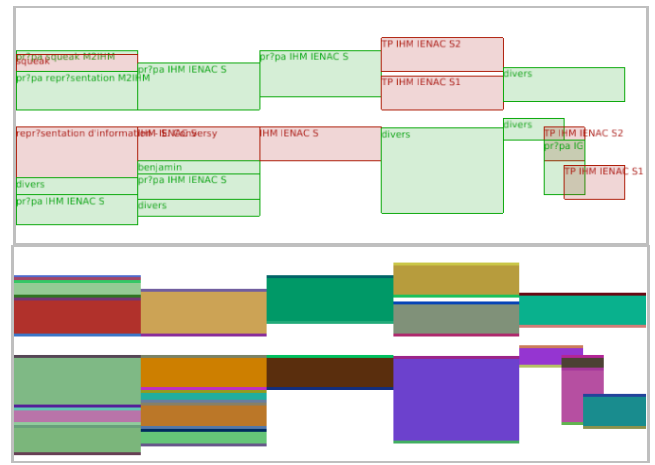

Figure 9. The "display" view (top) and the corresponding "picking" view (bottom) of a calendar. The picking algorithm uses unique colors for each picking object, which explains the colorful picking view.

### A. Interaction specification

Again, a traditional algorithm uses the positions and analytical distance computation to decide the reaction to user events.

*1) Description*

With MDPC, the "Display" view of each calendar event is a rectangle (Figure 9., top). The top side reflects the date and time of the start of the event, while the bottom side reflects the date and time of the end of the event. The width of the event is not tied to the data: it is equal to the width of a column, here a seventh of the window (since a week contains seven day). When multiple calendar events overlap, the corresponding rectangles share the column width.

The picking view of each event is composed of three juxtaposed rectangles (Figure 9., bottom). The middle rectangle is similar to the rectangle of the display view, and its height depends on the event duration. A Drag'n'Drop of this rectangle allows modifying both the start and end time without modifying its duration. The two other rectangles allow the user to pick the top (resp. bottom side) of a calendar event, and change by direct manipulation the start (resp. end) of the event. The modification of the data is done thanks to an inverse transformation, as explained in the next section.

*2) Simplicity*

The gain in simplicity is the same as the previous examples: this improves Closeness of Mapping and Visibility.

B. *Implementation*

The previous examples use Java and a scene-graph. They illustrate the use of picking views for managing interaction state, and for avoiding analytical computation. We implemented the calendar example with Tcl [13] and OpenGL [12], and by relying on a data-flow. This demonstrates not only the use of picking views, but also the use of inverse transformations, the second principle of MDPC. It also shows that MDPC is independent from the language, and that MDPC does not require a scene-graph.

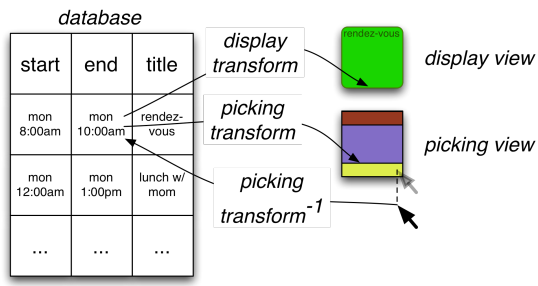

Figure 10. Display and Picking view of a calendar event. The position of the cursor is transformed back to the conceptual model by using the inverse picking transform.

*1) Description*

The architecture is illustrated in (Figure 10. ). Calendar events are stored in a relational database table. The table includes a "start", an "end" and a "title" column. A SQL select allows selecting visible events and computing the value needed for the visualization.

Each frame rendering triggers two OpenGL-based redisplay functions, one for the display view (proc view, display view, Figure 12. ), and one for the picking view (proc view, picking view, Figure 12. ). The display transformation fills pixels in the frame buffer, while the picking transformation fills pixels in an offscreen buffer. Both transformations share a *transf* function (Figure 12. , middle-left). *transf* first wraps the data multiple times on X and Y (Figure 11. ). The *wrap* function (shown in Figure 12. , bottom-left) is more complex than necessary (since we only use the week view), but serves as a demonstration that even a complex function can be reversed. Once wrapping is done, the position in the day is computed, and displayed on the screen's Y dimension quantitatively (yInDay). This leads to a 2-D position expressed in terms of cells (e.g. (3; 4.5)), which is then multiplied by the actual display size of a cell (CellWidth x CellHeight). Finally, the *transf* function applies a user-controlled pan and zoom. A final computation shifts the x position of events inside a cell to take into account parallel events (Figure 9. , right).

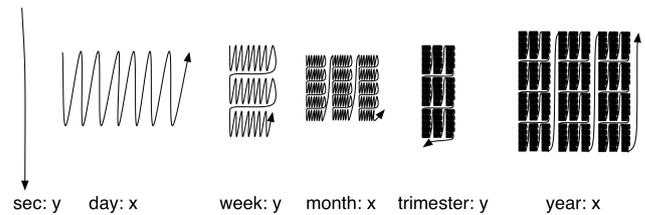

Figure 11. A calendar is a wrapped view of time over X and Y

The code that manages user input is shown in Figure 12. right. When the user presses on and moves one of the small rectangles in the picking view of a calendar event, an inverse transformation is applied on the X and Y dimensions of the Move event. Since the position of the rectangles is the result of the application of a continuous and monotonous function on a scalar (a time), it is sufficient to apply the inverse function to the position of the cursor to get the corresponding value in the referential of the data model. The inverse *transf* is shown in (Figure 12. , middle-right), and the inverse *wrap* is shown in (Figure 12. , bottom-right). Finally, a SQL query update modifies the data in the data table. After each modification (hence after each movement), the system triggers a redisplay, and the modification is visible immediately.

*2) Simplicity*

The display is the result of the application of a function on the data. The first advantage is that the understanding of how the model is transformed on the screen is easier to grasp, because it only depends on an identified flow, and is not spread around the entire program (Figure 12. ): in other words, spaghettis untangle [2]. This improves Locality [3] and thus Visibility. The second advantage is that if the function is a reversible transformation (which is the case here), the design of the function that transforms user manipulations into results on the model is straightforward: it consists in applying inverse sub-functions in reverse order. Moreover, the visualization of the text helps to design such an inverse function, because of the Symmetry [3] between transformation and their inverse (Figure 12. )When designing the display and the interaction, a good way for a designer to get confidence in the code is to target and reach this symmetry, and verify that at each sub-function corresponds an inverse sub-function.

Using functional code enables the implementation to use a data-flow. When applying modifications to the model, all depending variables (in particular all graphical positioning properties) are recomputed and displayed immediately. There is no need to manage consistency, which reduces the Viscosity cognitive dimension. Variables external to the model also benefit from data-flow. For example, the width and height of a cell depends on the the containing window. When the user resizes the window, the size of cells adapts "automatically".

*3) Performance*

If it is simpler to manage than analytical computation, this architecture is more costly in terms of computation. For example, it is necessary to recompute at each modification the tessellation and the rasterization of each graphical object. This behavior is similar to 3D applications and games: with 3D scenes, since the point of view may differ for each frame, coders do not bother implementing algorithms that manage damaged zones, and usually redisplay all objects. We have considered that given the computing power at disposal since the advent of 3D games, it is more beneficial to trade performances for ease of coding. Besides, the description with a data flow can help optimizing performances: it is possible to consider the chain of transformation from data to pixels as a compiler, and use automatic optimization provided by a graphical compiler [14] (partial evaluation, automatic cache, dead-code elimination, etc). Finally, if a data-flow may be more costly in terms of computation, it is less costly in terms of memory since it does not retain graphics.

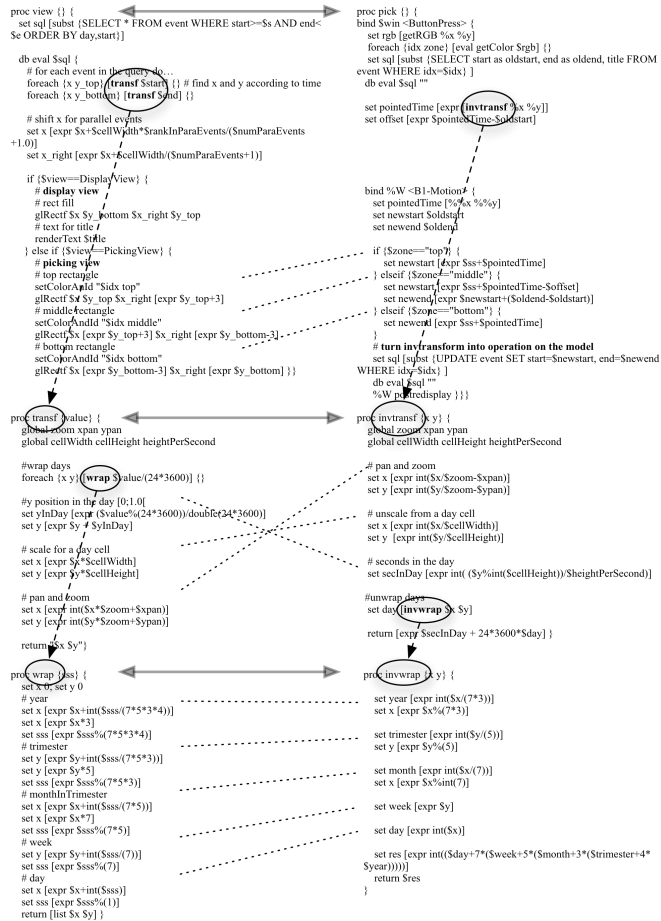

Figure 12. Actual code for calendar - Left: view (display & picking), transformation (transf), and wrapping (wrap) - Right: their inverse (pick, invtransf, invwrap). Note the symmetry or anti-symmetry of functions and their inverse.

## VII. DISCUSSION

This section synthesizes the benefits of using MDPC for specification and implementation.

### A. Software Engineering

As explained in [4], MDPC improves modularity of software. The role of the Controller of MDPC is limited to the management of the dynamics of the interaction state. In the Drag'n'Drop and Magnetic guide, the controller is reduced to the state-machine. In the Calendar example, the Controller is the interaction code. Since the Controller is independent from geometrical or layout transforms, it can be reused across multiple interactions. For example, if a pan is applied to the D'n'D or Magnetic guide scenes, there is no need to change the interaction code. This is particularly visible in the Calendar example: the same code can be used regardless of the fact that pan and zoom is handled by the application. One can add a rotation at the end of the transf function, and its inverse at the beginning of the invtransf function (for example to implement interactions from [15]), with no need to modify further the existing code. The interaction of the user will still be perfectly transformed into operations on the model.

It is important to note that it is the combination of picking view and inverse transformation that enables this feature. Using picking views radically simplifies the code, and avoids the need for complex adaptation of analytical code when one adds a new transformation. And transformations are an abstraction that is both independent from the notion of interactive state, and that can still be applied easily to the reification of interactive state into picking views.

### B. Implementation: scene graph considered harmful

The implementation of the calendar system uses a paradigm that contrasts with the paradigm relying on a scene graph. Often, implementers use a scene-graph to retain the properties of the graphical objects, and to optimize the rendering. In fact, a scene graph is a "cache" of the rendering pass. As a cache of graphical properties, it relieves the apparent obligation to retain the graphical objects for subsequent redisplay. As a cache of transforms, it optimizes the redisplay: often, the modification between two frames is minor, and one can expect better performance if previous computation is reused.

However, as with any "cache", consistency must be dealt with. Consistency management is known to be error-prone, and even if it seems compulsory to users of scene-graph, it requires care to be taken, hence time and resources, at the expense of other concerns. We think that graphics management, user input management, and data update are hindered by consistency management. The data-flow architecture inherently eliminates cache management problems, since there is no cache anymore.

As for performance of data-flow, we have already noticed that highly demanding 3D applications behave this way, and are efficient. Furthermore, some interaction requires drawing the entire scene. For example, resizing the window of the calendar application leads to a complete computation of all graphical elements in the scene. In this case, the advantage of the scene graph is null, since it does not act as a cache anymore (the cache is invalidated at each rendering pass). Besides, the use of a graphical compiler offloads optimization concerns from the programmer to a tool [14].

Finally, a number of services provided by a scene graph (ready-to-use graphical shape rendering, picking management) do not require a data structure that retains graphics. For example, the graphical properties need not be retained, since the transformations that lead to those graphical properties are retained in the code: graphical properties can be generated at each redisplay. In the same way, picking does not require a complex scene-graph. In the calendar example, picking is realized with a "pick by color" algorithm [16].

## VIII. RELATED WORK

A number of works have tackled usability of programming, including psychology of programming, cognitive dimensions of notation [9], or API usability [17]. For example, [10] and [11] enable the programmer to describe interactive state with state machines [18]. Most usability studies target general-purpose languages or APIs rather than tools for building interactive systems [3]. Exceptions include Myers' study of the programming practices of graphical designers [1]. Our work builds up on these concerns, and proposes a practical method that aims at improving usability of specification and implementation of graphical interaction. Artistic resizing is a technique that enables to specify how graphical components resize when users resize the container window [19]. It is an example of how a specification can be turned from a program to graphical description. Our work is in the same vein, in that it improves the Closeness of Mapping between the phenomenon and its description.

Describing graphics with Data Flow has been extensively studied in the past. For example, Fabrik is a direct manipulation - based user interface builder that enables a designer to specify transforms between widget with a visual flow language [20]. Events flow in the same flow graph that describes the geometrical transforms, so that they are automatically transformed to a position relative to the graphically transformed widget. Garnet uses one-way constraints, which can be considered as data flow, to propagate changes [21]. In order to improve interactive graphics programming, [22] proposes solutions to facilitate mixing of data flow of input and scene graph for output.

The inverse of model-view matrix is often used to retrieve an object that has undergone multiple 3D transforms (due to change of point of view, or due to modeling) [12]. [23] discusses how to enable users to change data through visualization and a data-flow. Metisse [15] and Façade [24] rely on inverse transforms to handle user manipulation in rotated views. However, none discusses how to design inverse transformations to reflect users' manipulation into the models.

## IX. CONCLUSION

We have presented how the MDPC architectural pattern, based on picking views and inverse transformations, can help at specifying and implementing graphical interaction. We have evaluated positively its ability at describing a large range of graphical interaction. We have also assessed the simplicity of description, by identifying the benefits (modularity, closeness of mapping, visibility, locality and symmetry of code). Of course, there are some drawbacks (terseness and performances in certain cases), and the claims, even if supported analytically, must be experimentally tested. Furthermore, we do not claim that MDPC is adapted to all graphical interaction. For example, one would better apply a modulo operation to cursor position to align objects on a grid, instead of relying on one picking shape per row or column on the grid. However, we think that thinking in terms of reified spatial mode of interaction and transforms helps at designing an interaction. In the future, we plan to separate even further the implementation of graphics and the implementation of transformation by using specialized languages (e.g. SVG as in [14]), and to explore optimization and especially cache management.